\documentclass{emulateapj}
\def\sun{\ifmmode\odot\else$\odot$\fi}

\input{psfig.sty}
\shorttitle{The host galaxies and black holes of typical $z \sim 0.5-1.4$ AGN}
\shortauthors{A. Alonso-Herrero et al.}

\begin{document}

\title{The Host Galaxies and Black Holes of
Typical $ z \sim 0.5-1.4$ AGN}

\author{Almudena Alonso-Herrero\altaffilmark{1,2}, Pablo G. 
P\'erez-Gonz\'alez\altaffilmark{3,2}, George H. Rieke\altaffilmark{2},
David M. Alexander\altaffilmark{4}, Jane R. 
Rigby\altaffilmark{5}, Casey Papovich\altaffilmark{2}, 
Jennifer L. Donley\altaffilmark{2}, 
and 
Dimitra Rigopoulou\altaffilmark{6} }

\altaffiltext{1}{Departamento de Astrof\'{\i}sica Molecular 
e Infrarroja, Instituto de Estructura de la Materia, CSIC, 
E-28006 Madrid, Spain; E-mail: aalonso@damir.iem.csic.es}
\altaffiltext{2}{Steward Observatory, University of Arizona, Tucson,  AZ 85721}
\altaffiltext{3}{Departamento de Astrof\'{\i}sica y Ciencias de 
la Atm\'osfera, Universidad Complutense de Madrid, E-28040 Madrid, Spain}
\altaffiltext{4}{Department of Physics, University of Durham, Durham, DH1 3LE,
UK}
\altaffiltext{5}{Observatories of the Carnegie Institution of Washington,
  Pasadena, CA 91101}
\altaffiltext{6}{Nuclear Physics and Astrophysics, Keble Road, Oxford OX1 3RH,
 UK}

\begin{abstract}
We study the  stellar  and star formation properties of the host galaxies of
58 X-ray selected AGN in the GOODS portion of the 
{\it Chandra} Deep Field South (CDF-S) region
at $z\sim 0.5-1.4$. The AGN are selected such that their 
rest-frame UV to near-infrared
spectral energy distributions (SEDs) are dominated by stellar emission, i.e.,
they show a prominent $1.6\,\mu$m bump, thus minimizing the AGN emission 
'contamination'. 
This AGN population comprises approximately 50\% of the X-ray selected 
AGN  at these redshifts. 
Using models of stellar and dust
emission we model their SEDs to derive  stellar masses 
($\mathcal{M}_*$) and total (UV+IR) star formation
rates (SFR). We find that  AGN 
reside in the most massive galaxies at the redshifts probed
here. Their characteristic stellar masses 
($\mathcal{M}_* \sim  7.8 \times 10^{10}\,{\rm M}_\odot$ 
 and $\mathcal{M}_*  \sim 1.2 \times 10^{11}\,{\rm M}_\odot$ 
at  median redshifts of 0.67 
and 1.07, respectively) appear to be 
representative of the X-ray selected AGN population
at these redshifts, and are intermediate between those
of local type 2 AGN and high redshift ($z \sim 2$) AGN. 
The inferred black hole masses ($\mathcal{M}_{\rm BH} \sim 2 \times
10^{8}\,{\rm M}_\odot$) of typical AGN  are similar 
to those of optically identified quasars at similar redshifts. 
Since the AGN in our sample are much less luminous ($L_{\rm 2-10keV}
<10^{44}\,{\rm erg \, s}^{-1}$) than quasars, typical AGN have 
low Eddington ratios ($\eta 
\sim 0.01-0.001$). 
This suggests that, at least at intermediate redshifts, 
the cosmic AGN 'downsizing' is due to both a 
decrease in the characteristic stellar mass of typical host galaxies,  and less
efficient accretion.
Finally there is no strong evidence in AGN host galaxies 
for either highly suppressed star
formation (expected if AGN played a role in quenching star formation) or
elevated star formation when compared to mass selected (i.e., IRAC-selected) 
galaxies of similar 
stellar masses and redshifts. This may be explained by the fact that 
galaxies with $\mathcal{M}_*  \sim 5\times 
10^{10}-5 \times 10^{11}\,{\rm M}_\odot$ are still being assembled at the redshifts probed
here.

\end{abstract}

\keywords{galaxies: active --- galaxies: evolution  --- galaxies:
  high-redshift --- galaxies: stellar content --- infrared: galaxies}

\section{Introduction}

One of the challenges faced by 
galaxy formation models is to explain the population of 
today's red massive quiescent elliptical galaxies. In the current
hierarchical galaxy formation paradigm massive galaxies are formed via
mergers of less massive galaxies which in turn fuel intense star
formation and feed massive black holes. One of the main difficulties is
to find mechanisms to stop the processes of intense star formation, 
 and to allow galaxies to migrate from the so-called 
'blue cloud' or late-type
star forming galaxies  to the so-called 'red
sequence' or early-type quiescent galaxies 
(see e.g., Bell et al. 2004, 2007).  Feedback from active galactic  
nuclei (AGN) has been proposed as an efficient process for suppressing any
further star formation in the late stages of galaxy evolution, while 
still allowing for continuing black hole growth (see e.g., Springel, di
Matteo, \& Hernquist 2005 and Croton et al. 2006). See Hopkins et al. 
(2007, and references
therein) for a detailed discussion on this and other related issues.

Some tantalizing evidence of the possible role of AGN in galaxy evolution is 
the location of local optically selected AGN (e.g., Salim et al. 2007; Martin
et al. 2007) and 
moderate-$z$ X-ray selected AGN (S\'anchez et al. 2004;
Nandra et al. 2007) in the transition between 
the 'red sequence' and the top of the 'blue cloud', the region also known as the 
'green valley'.   These intermediate colors 
may indicate that AGN play a role in causing or maintaining the quenching of
star formation.  However, 
in the local universe AGN with strongly 
accreting black holes tend to be hosted in massive galaxies with blue 
(i.e., star-forming) disks and young bulges (Kauffmann et al. 2003b, 2007)
implying a close link between the growth of black holes and bulges. 
Clearly the relationship between AGN and star-formation is a matter of strong
debate. 

 About half of the sources with X-ray
luminosities $\gtrsim 10^{41}\,{\rm erg \, s}^{-1}$ (i.e., suggestive of the
presence of a moderately luminous AGN) detected in deep ($\ge 1\,$Ms) X-ray
surveys  do not show 
broad lines or high 
excitation lines characteristic of AGN in their 
optical spectra (see e.g., Barger et al. 2001;
Cohen 2003; Szokoly et al. 2004
and review by Brandt \& Hasinger 2005). 
Since the AGN emission of these  optically-dull 
AGN does not dominate 
their rest-frame UV to near-infrared (NIR) emission (Rigby et al. 2006), 
they are the ideal targets to study
their host galaxies and investigate the
role of AGN in galaxy evolution. 

In this paper we study the host galaxies of X-ray selected 
AGN with stellar dominated spectral energy distributions (SEDs) at
intermediate redshifts ($0.5<z<1.4$) in the  
{\it Chandra} Deep Field South (CDF-S) using UV, optical, NIR, 
and {\it Spitzer} data. The AGN host galaxy properties are 
then compared with those of IRAC-selected (i.e., stellar mass
selected) galaxies at similar distances studied by 
P\'erez-Gonz\'alez et al. (2008). 
Throughout this work we assumed the following
cosmology: $H_0=70\,{\rm km \, s}^{-1} \, {\rm
  Mpc}^{-1}$, $\Omega_{\rm M=}0.3$ and  
$\Omega_\Lambda=0.7$.

\begin{figure*}
\includegraphics[width=6cm,angle=-90]{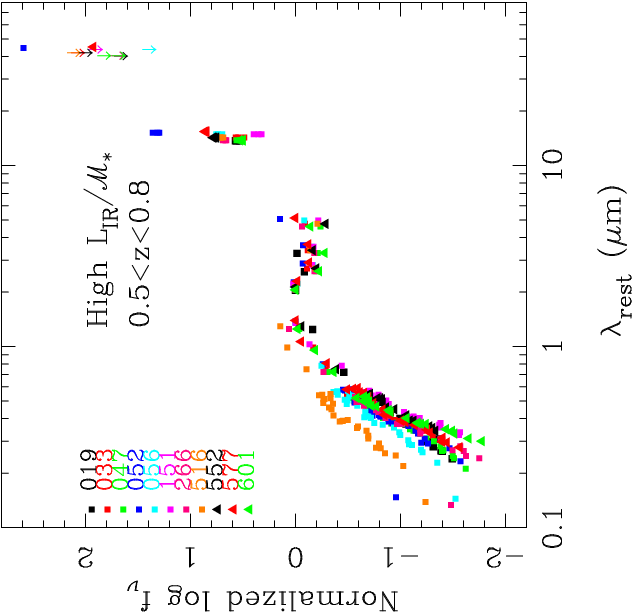}
\includegraphics[width=6cm,angle=-90]{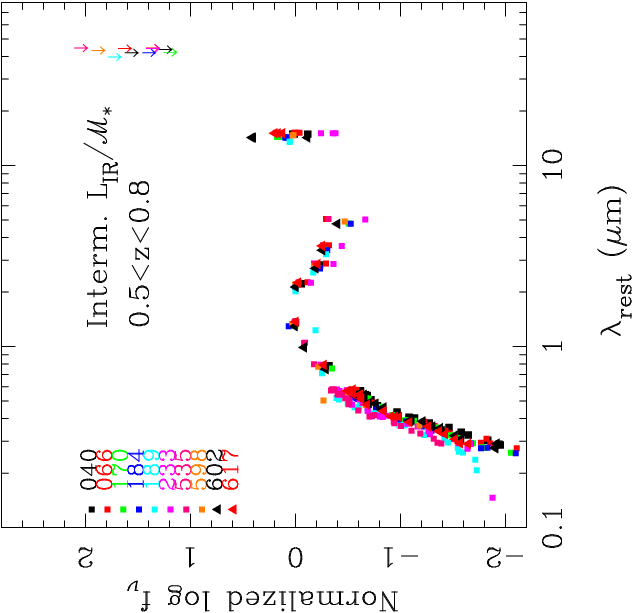}
\includegraphics[width=6cm,angle=-90]{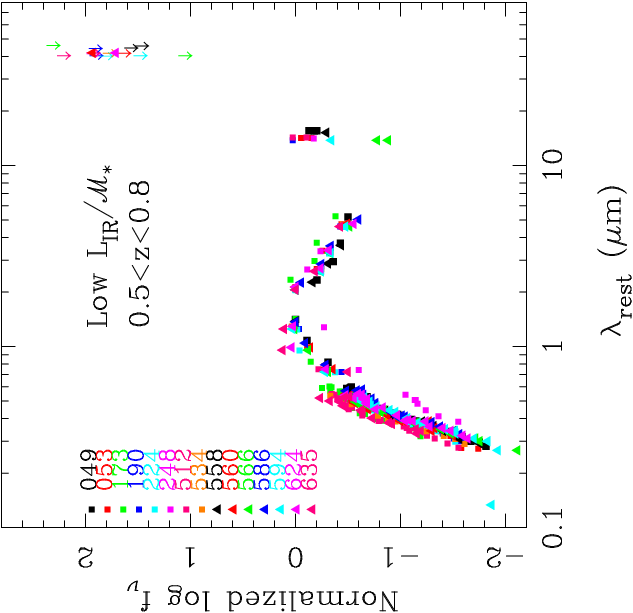}

\includegraphics[width=6cm,angle=-90]{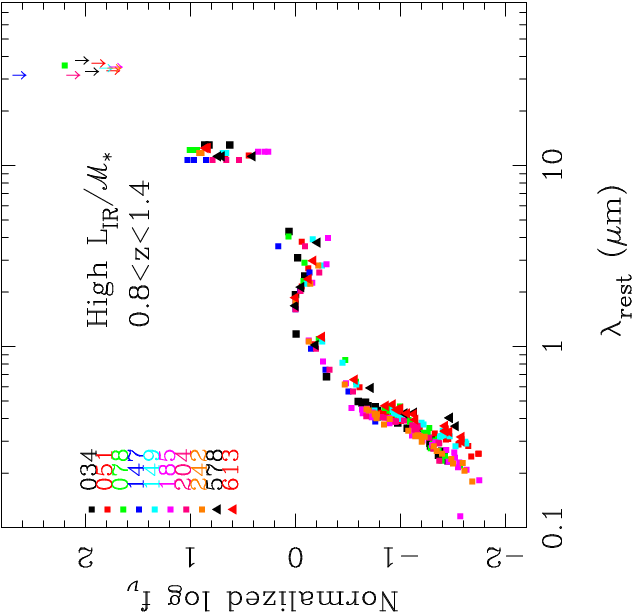}
\includegraphics[width=6cm,angle=-90]{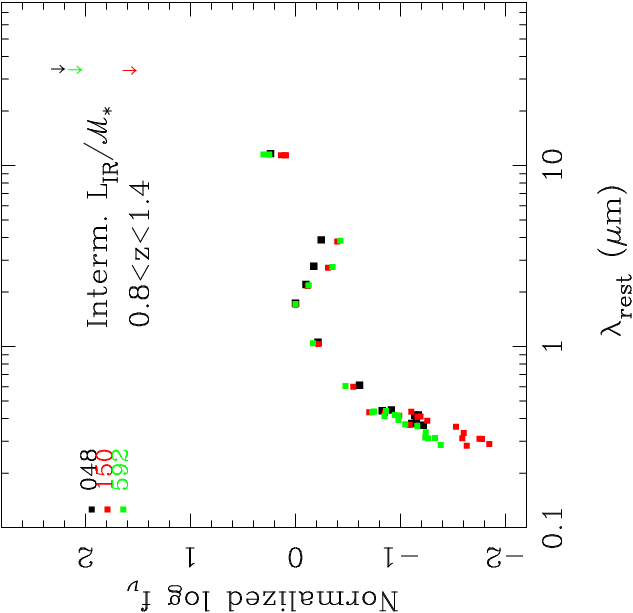}
\includegraphics[width=6cm,angle=-90]{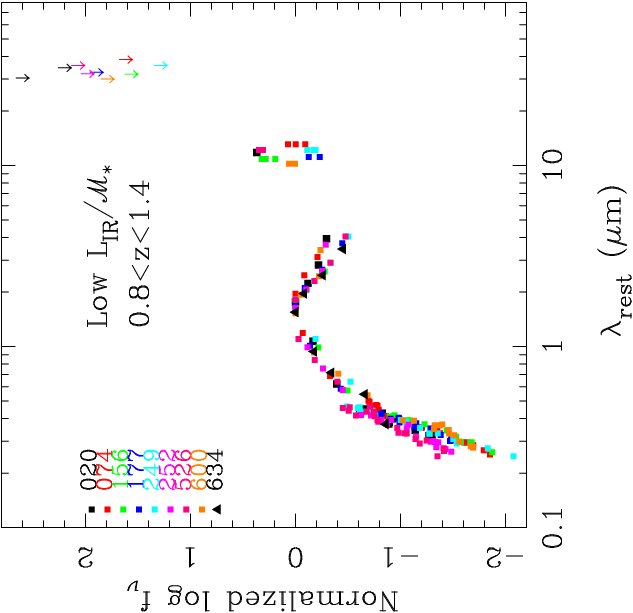}

%for preprint format
%\includegraphics[width=5.5cm,angle=-90]{figure1a.ps}
%\includegraphics[width=5.5cm,angle=-90]{figure1b.ps}
%\includegraphics[width=5.5cm,angle=-90]{figure1c.ps}

%\includegraphics[width=5.5cm,angle=-90]{figure1d.ps}
%\includegraphics[width=5.5cm,angle=-90]{figure1e.ps}
%\includegraphics[width=5.5cm,angle=-90]{figure1f.ps}

\caption{{\it Upper panels:} 
Rest-frame SEDs normalized at $\sim 1-2\,\mu$m of CDF-S
  AGN 
at $0.5<z<0.8$ whose integrated UV-through-NIR
SEDs are dominated by stellar emission. The numbers given next to
the symbols in the upper left corners of the plots are the Giacconi et
al. (2002) IDs. Left to right galaxies are plotted for decreasing IR
luminosity to stellar mass  
ratios relative to the mass dependence found for IRAC-selected galaxies (see
more details in \S5.2). These ratios are a measure of the specific SFR 
if the IR luminosity
is not dominated by AGN emission. {\it Lower panels:} Same but for
  AGN at $0.8 < z < 1.4$.}
\end{figure*}

\section{Sample Selection and Data}
We first started with all the CDF-S X-ray sources 
(Giacconi et al. 2002; Alexander et al. 2003) with spectroscopic and 
photometric redshifts in the range of $0.5<z<1.4$ (Zheng et al. 2004). 
Then we restricted ourselves to  those X-ray sources in the 
Great Observatories Origins Deep Survey (GOODS) 
portion of the CDF-S region which has the deepest
{\it Spitzer} IRAC (Fazio et al. 2004) and 
MIPS (Rieke et al. 2004) observations (see P\'erez-Gonz\'alez et al. 
2008).  We cross-correlated the positions of the X-ray sources
with the IRAC (simultaneous detections at $3.6$ and $4.5\,\mu$m)-selected 
galaxies of P\'erez-Gonz\'alez et al. (2008) using a separation of $ \le
1.5$arcsec.  The 75\% completeness limits for
the CDF-S field in this  
sample are $1.6\,\mu$Jy and $1.4\,\mu$Jy at 
$3.5\,\mu$m  and $4.5\,\mu$m, respectively. 

We also used the CDF-S 
photometric catalogs of P\'erez-Gonz\'alez  et al. (2008) to construct
the SEDs of the X-ray sources.   
These catalogs include the two other IRAC bands, UV, optical, NIR, 
and {\it
  Spitzer}/MIPS $24\,\mu$m data (see P\'erez-Gonz\'alez et al. 
2005, 2008 for a complete description of the dataset and source matching).
The MIPS $24\,\mu$m catalog is 75\% complete down to $80\,\mu$Jy. 
We finally 
cross-correlated the X-ray sources  with the {\it
  Spitzer}/MIPS $70\,\mu$m catalog for the CDF-S  
 (Papovich et al. 2007) 
which is $50\%$ complete for sources with flux densities down to 
$f_\nu (70\mu{\rm m}) \sim 3.9\,$mJy. 

Out of the 112 X-ray sources with IRAC detections in the field 
described above, 
we selected for this study AGN with stellar-dominated UV through NIR SEDs, 
and in particular with a strong $1.6\,\mu$m bump.  
Our selection  thus excluded 
X-ray sources with AGN-dominated SEDs such as IR
power-law galaxies (Alonso-Herrero et al. 2006; Donley et al. 2007), 
IRAC color-color selected AGN (Lacy et al. 2004; Stern et al. 2005)  and 
galaxies without a prominent $1.6\,\mu$m bump (e.g., Daddi et al. 2007). 
We also removed from our sample 
X-ray sources with nearby companions of similar brightness 
which could contaminate
the observed mid-IR (MIR) SEDs, in particular the IRAC bands. 
The final sample contains 58 AGN.

Most (52 of the 58)  AGN in our sample  have spectroscopic redshifts 
and type classifications (Szokoly et al. 2004; Vanzella et al. 2006). The
majority (46 out of the 52) are classified as optically-dull, that is, 
AGN without any evidence for accretion from their
optical spectra. These included AGN classified as low excitation  and
absorption line AGN (see Szokoly et al. 2004). 
The remaining 6 AGN with spectroscopic information show 
high excitation lines or broad lines (Szokoly et al. 2004),  
and will be referred to as
optically-active AGN. For the 6 AGN in our sample without spectroscopic
information we estimated photometric redshifts (see next section). 

The rest-frame SEDs of the 58 AGN are shown in Figure~1 for two
different redshift bins: $0.5<z<0.8$ and $0.8 < z < 1.4$, which 
correspond to approximately similar ranges of cosmic times for the assumed
cosmology.  

The selected AGN
have rest-frame absorption-corrected hard  ($2-10\,$keV) X-ray 
luminosities above $10^{41}\,{\rm erg\,s}^{-1}$ (from Tozzi et
al. 2006).  Figure~2 compares the X-ray column densities 
with the absorption-corrected hard 
X-ray luminosities (from Tozzi et al. 2006) for  all the X-ray sources
(open circles) detected by IRAC in
the GOODS field. We marked the AGN with
stellar-dominated SEDs 
selected for this study as  filled circles. 

Fig.~2 clearly shows that a large fraction
of AGN with hard X-ray luminosities below 
$\sim 10^{43}\,{\rm erg\,s}^{-1}$ 
have SEDs dominated by stellar emission (see also Donley et al. 2007). At
higher hard X-ray luminosities only those AGN with large X-ray column
densities have stellar SEDs (see also, Polletta et al. 2006, 2007). At 
low absorptions ($N_{\rm H} < 10^{22}\,{\rm cm}^{-2}$)  the 
AGN emission becomes more apparent in the integrated SEDs, but there are some
X-ray sources with low $N_{\rm H}$ with  stellar-dominated 
SEDs. These sources will be
studied in more detail in \S4.1. 

The X-ray properties of our sample of AGN (Fig.~2) 
are consistent with recent findings on the nature of optically-dull AGN. 
Rigby et al. (2006) estimated that in $\sim 50\%$ of optically-dull AGN    
the AGN emission lines could be 
diluted by the stellar emission from the host galaxy (see also Moran, Filippenko, 
\& Chornock 2002), whereas in the rest extinction from the host galaxy may  
be responsible for hiding the AGN optical lines. The latter conclusion was
based on the comparison of  the
inclination angle distributions of the host galaxies of optically-dull and
optically-active AGN. Recently, Caccianiga et
al. (2007) proposed that the most likely explanation for optical dullness at
low X-ray luminosities is
dilution by a massive host galaxy, while at high X-ray luminosities the 
dullness is due to absorption. 

\begin{figure}

 \includegraphics[width=8cm,angle=-90]{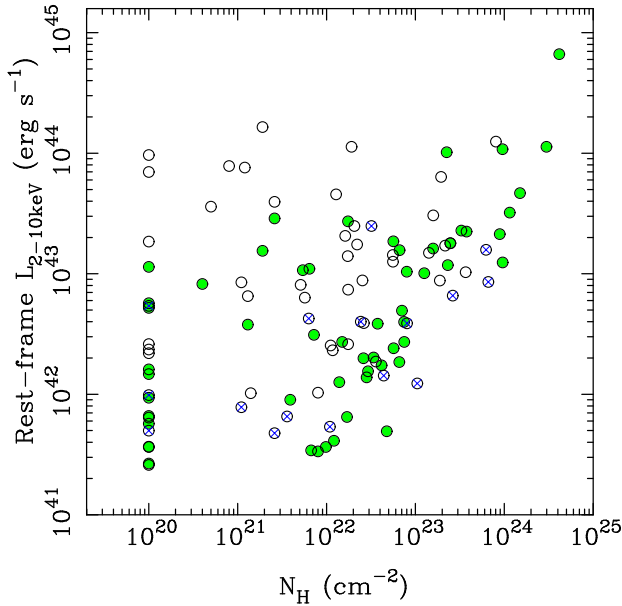}

%for preprint format

%\hspace{2cm}
% \includegraphics[width=10cm,angle=-90]{figure2.ps}
\caption{X-ray column density versus rest-frame absorption corrected hard
X-ray luminosity (both from Tozzi et al. 2006) for X-ray selected AGN (all circles) 
in GOODS-south detected by {\it Spitzer} at $0.5<z<1.4$. Our sample of AGN
with stellar-dominated SEDs are 
shown as filled circles. We also mark optically-dull AGN with nearby  companions
 not included in the present analysis. When
the X-ray column densities were estimated by Tozzi et al. (2006) to
be zero we plot them at $10^{20}\,{\rm cm}^{-2}$.}
\end{figure}

\section{Modelling of the stellar emission 
at $\lambda_{\rm rest}< 4\,\mu$m}

This section describes briefly (see P\'erez-Gonz\'alez et al. 2008 for a full
description) the procedure for modelling the SEDs at 
wavelengths $\lambda_{\rm rest}< 4\,\mu$m. This modelling is general to 
the IRAC-selected sample of P\'erez-Gonz\'alez et al. (2008) from which we 
extracted the stellar masses (this section) and SFRs (next section) 
for  our sample of  
AGN. The method involves a two
step process in which the galaxies with spectroscopic redshifts are fitted
first, and then used as templates to fit the SEDs 
(and photometric redshifts) of
those galaxies without spectroscopic redshifts. 

For  galaxies with spectroscopic redshifts 
the stellar emission is generated with the PEGASE code 
(Fioc \& Rocca-Volmerange 1997) assuming
a Salpeter IMF (Salpeter 1955) between 0.1 and 100\,M$_\odot$.
We assumed that the stellar emission of the 
galaxies can be described with one or two stellar
populations. In the case of the one stellar population model 
the star formation rate 
(SFR) is  modeled with 
a declining exponential (${\rm SFR} (t) \propto \exp^{-t/\tau}$). The 
four free parameters to fit are: extinction (using the Calzetti et al. 2000
law),
metallicity, the time scale of the exponential law ($\tau$), and the 
age ($t$) of the stellar population. In the case of two stellar 
populations, the old stellar
population is described as for the one stellar 
population model (four free parameters), and the young
stellar population is assumed to have been formed in an instantaneous burst
with three free parameters: extinction, metallicity, and age. 
For the two stellar population models the burst strength 
is another free parameter that relates the mass of young stars with
the total mass of the galaxy. In addition to the stellar emission, we include
the hydrogen gas emission in the form of nebular continuum and emission
lines.

Once the galaxies with spectroscopic redshifts are fitted, they are used as
templates for fitting the SEDs of 
those galaxies without a spectroscopic redshift.
With the best model and best photometric redshift established, the stellar
mass of the galaxy is obtained by scaling the models to the observed SED. The
value of the stellar mass ($\mathcal{M}_*$) 
is computed as the average of the stellar masses
computed for each observed photometric band. 
Although most (52 out of 58) 
of the AGN in
our sample have spectroscopic redshifts (Szokoly et al. 2004 and 
Vanzella et al. 2006), we refitted their SEDs using the library of
templates. The stellar masses derived  with the photometric redshift 
were then rescaled to the spectroscopic
redshift.

As discussed in great detail by P\'erez-Gonz\'alez et al. (2008) there are a
variety of effects that can introduce systematics in the determination of
stellar masses. In addition to the intrinsic uncertainties
associated with fitting the SEDs (typically a factor of $2-3$), other effects
include the choice of stellar population libraries, 
extinction law, and IMF, as well as the  
number of stellar populations (in our case, one
population vs. two populations). These effects introduce uncertainties of the
order of or smaller than those intrinsic to the SED fitting. 
%Since for most of the galaxies in the
%study of P\'erez-Gonz\'alez et al. (2007) there are no spectroscopic 
%redshifts available, the SEDs of these galaxies are fitted using the templates
%generated for the spectroscopic sample, and in addition to the parameters
%given above the photometric redshift of the galaxy is also fitted
%using a library of templates generated for galaxies with spectroscopic
%redshifts. 

Even though we selected our AGN such that their SEDs are clearly 
 dominated by stellar emission, 
a valid concern is the possible effect of the AGN emission in the
 determination of the stellar masses. In particular, such concern was
 raised by Daddi et al. (2007) when estimating the stellar masses
of $z\sim 2$ galaxies with MIR excesses thought to host obscured
 AGN. These authors argued that the possible AGN contribution in
 the form of hot dust at $\lambda_{\rm rest} \ge 
1.6\,\mu$m 
might cause overestimation of the stellar masses, but only at a modest level
and generally within the systematic errors.

Daddi et al. (2007) estimated the stellar masses using empirical 
calibrations  based on $B$, $z$, and $K$-band photometry. We believe our 
method is less susceptible to the effects of AGN because we use 
population synthesis models to fit
 the entire observed SEDs, and the stellar masses are computed 
as the average of
 the stellar masses in all the photometric points.  
P\'erez-Gonz\'alez et al. (2008) discussed
 in detail the effects of the presence of a hot dust component (which could be
 associated
 with an AGN) in the determination of the stellar masses of AGN. They
 concluded that the AGN effects on the estimated stellar masses 
are negligible for X-ray sources with observed
 (i.e., not corrected for absorption) luminosities $L_{\rm X} < 10^{44}\,{\rm
 erg\,s}^{-1}$ (see also Donley et al. 2007). 
All the AGN in our sample are well below this limit.

\section{The IR emission}

In this section, we evaluate whether the MIR emission of 
our sample of AGN is likely to be due to the AGN or to star formation. 
We then use the {\it Spitzer} data to model the total
IR ($8-1000\,\mu$m) luminosity ($L_{\rm IR}$) and to derive SFRs.   

\subsection{The observed $24\,\mu$m emission of low $N_{\rm H}$ AGN}

The X-ray column density can be used to classify AGN into type 1 (direct view
of the AGN, $N_{\rm H}<10^{22}\,{\rm
  cm}^{-2}$) and type 2 (obscured view of the AGN). 
Rigby et al. (2004) using  local AGN 
templates showed
that for low column densities the $24\,\mu$m to hard X-ray flux ratio should
remain approximately constant, whereas for high column densities this ratio
varies by factors of a few. We  test whether 
the $24\,\mu$m emission can be solely explained as produced by dust heated by
the putative AGN, or whether an additional mechanism (i.e., star formation) is
needed. This is tested for the  eighteen 
optically-dull AGN (that is, those without evidence for accretion from the
optical spectra) in our sample with  $N_{\rm H}<10^{22}\,{\rm
  cm}^{-2}$. To do so, we scaled the 
median quasar template of Elvis et al. (1994) to 
the $2-10\,$keV flux (from Tozzi et al. 2006) 
for each of the 18 optically-dull AGN with low $N_{\rm H}$. 
The predicted AGN 
$24\,\mu$m flux densities were then compared with observed values.

We only find five optically-dull AGN with low $N_{\rm H}$ 
(those in the left panel of Fig.~3)
for which between 30 and 100\% of their observed 
$24\,\mu$m flux densities could be accounted for with the predictions from 
their AGN hard X-ray fluxes. These five AGN show the highest X-ray fluxes
among the optically-dull AGN with low $N_{\rm H}$. 
For the rest (including some galaxies with $\lambda_{\rm rest} \sim 
3-5\,\mu$m excesses), 
the predicted AGN $24\,\mu$m flux 
densities are  below  approximately 
15\% of the observed value, and thus most of their MIR
emission is probably produced by star formation. 
In our analysis, we will assume that star formation dominates the $24\,\mu$m
emission of these galaxies and of those that are similar except for larger
absorbing columns for their X-ray sources.

For comparison the middle and
right panels of Fig.~3 show a few examples of SEDs of optically-active AGN
with $N_{\rm H} < 10^{21}\,{\rm cm}^{-2}$, 
and their corresponding scaled quasar templates. The middle panel includes
three optically-active AGN in our sample (Szokoly 52, 53, and 78). Only the
MIR emission of one of them (Szokoly 53) is fully accounted 
for by the predictions from
the AGN X-ray flux.  From the right panel of Fig.~3, it  is clear that the AGN signatures 
 (UV bump and hot dust emission around $\lambda_{\rm
  rest} \sim 2-5\,\mu$m and beyond) become dominant for
the most X-ray luminous AGN (see e.g., Barmby et al. 2006; Polletta et al. 
2006, 2007; Donley et al. 2007).

%Summarizing, most of the MIR
%emission of optically-dull AGN with low $N_{\rm H}$  
%is likely to be due to star formation with a 
%relatively small
%contribution arising from the AGN. Also their 
%the optical ``dullness'' is a
%combination of the intrinsically less luminous AGN and dilution by the
%emission from the host galaxy and star formation. 

\begin{figure*}

\includegraphics[width=6cm,angle=-90]{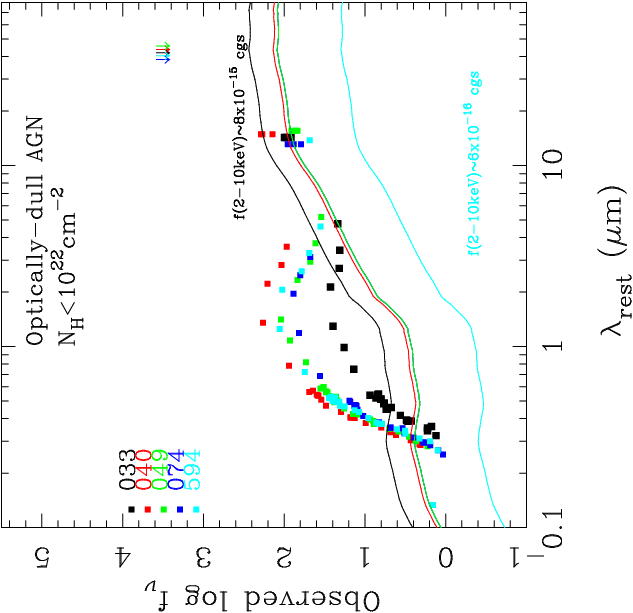}
\includegraphics[width=6cm,angle=-90]{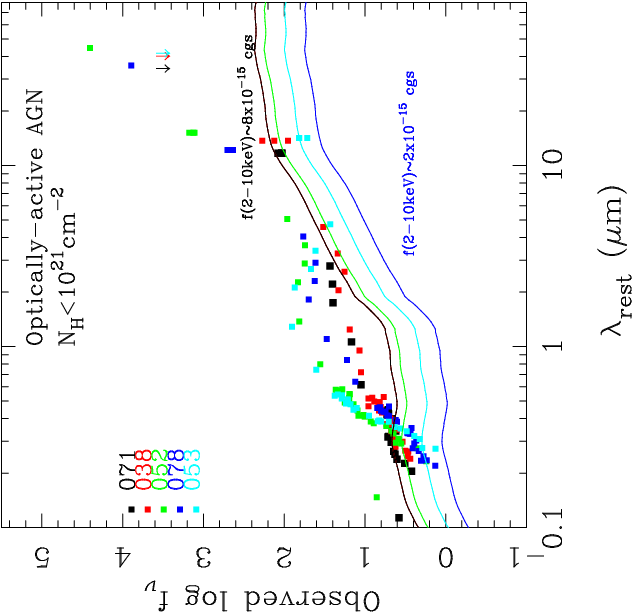}
\includegraphics[width=6cm,angle=-90]{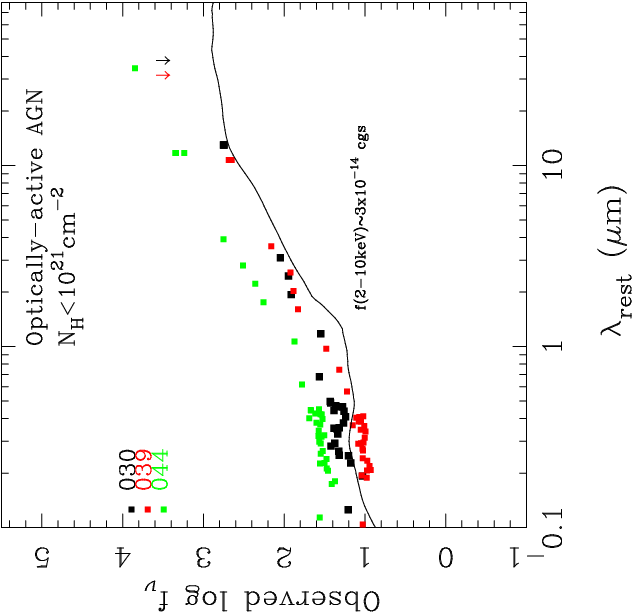}

%for preprint
%\includegraphics[width=5.5cm,angle=-90]{figure3a.ps}
%\includegraphics[width=5.5cm,angle=-90]{figure3b.ps}
%\includegraphics[width=5.5cm,angle=-90]{figure3c.ps}

\caption{{\it Left panel:} Examples of rest-frame observed (in $\mu$Jy) 
SEDs (filled squares) of optically-dull AGN with low X-ray absorptions
  ($N_{\rm H}< 10^{22}\,{\rm cm}^{-2}$). For each galaxy the Elvis et al. 
(1994) median quasar template (solid line in the same color as the galaxy data
points) is plotted scaled to the hard X-ray flux. 
These are those optically-dull AGN with low $N_{\rm H}$
where the predicted AGN emission accounts for
between approximately $30\%$ and $100\%$ of the observed $24\,\mu$m flux
density. {\it
  Middle and Right panels:} Examples of optically-active AGN in the same
redshift interval studied here with low 
X-ray absorptions ($N_{\rm H}< 10^{21}\,{\rm cm}^{-2}$). In the middle panel,
three AGN (Szokoly 52, 53, and 78) are included in our sample of AGN. Again the Elvis et
al. (1994) template is scaled to the observed hard X-ray flux of each AGN.}
\end{figure*}

\subsection{Modelling of the IR emission}

After the stellar emission was modeled, as described in \S3,  
the predicted stellar fluxes were
subtracted from the observed photometric data points at $\lambda_{\rm rest} >
4\,\mu$m. 
The resulting dust emission out to the MIPS $24\,\mu$m photometric point 
was then fitted using the   
Chary \& Elbaz (2001) models to derive the IR 
luminosity.  For the  $24\,\mu$m non-detections (10 out of the 58 AGN in our sample), 
the $L_{\rm IR}$  was computed assuming an upper limit to
the $24\,\mu$m flux density of $60\,\mu$Jy, which is the 50\% completeness limit of our
catalog. The unobscured star formation is assumed to be  
traced by the UV monochromatic 2800\AA \ luminosity, as fitted by the stellar
model.  The total SFRs were computed as the sum of the IR 
and UV luminosities converted to SFRs using the prescriptions of
Kennicutt (1998).

Although we do not use the MIPS $70\,\mu$m photometric points to model the IR
luminosities, we can check if the 
$f_\nu(70\,\mu{\rm
  m})/f_\nu(24\,\mu{\rm m})$ ratios or upper limits 
are consistent with star formation, as assumed. 
Only 3  optically-dull AGN are  
detected at $70\,\mu$m, all of them in the $0.5 < z < 0.8$ bin (see Papovich et
al. 2007 for more details). 
None of these three sources with $70\,\mu$m detections 
have observed $f_\nu(70\,\mu{\rm m})/f_\nu(24\,\mu{\rm m})$ ratios
consistent with those expected from hot dust arising from 
a $\nu f_\nu=$ constant distribution 
($f_\nu(70\,\mu{\rm m})/f_\nu(24\,\mu{\rm m})=2.9$, similar to optically
selected quasars, Elvis et al. 1994 see Fig.~3, or AGN-dominated SED, see e.g.,
Alonso-Herrero et al. 2006). Neither are these colors consistent with those of
IR bright AGN such as Mrk~231 
($f_\nu(70\,\mu{\rm m})/f_\nu(24\,\mu{\rm m}) \sim 7-9$, for our redshift
range). The behavior of the $f_\nu(70\,\mu{\rm m})/f_\nu(24\,\mu{\rm m})$ ratio
of optically-dull AGN is however similar to other
sources (both detected and undetected in X-rays) 
in the same redshift range (see Papovich et al. 2007), as well as to 
empirical templates of local star-forming galaxies.

\begin{figure}
 \includegraphics[width=4.5cm,angle=-90]{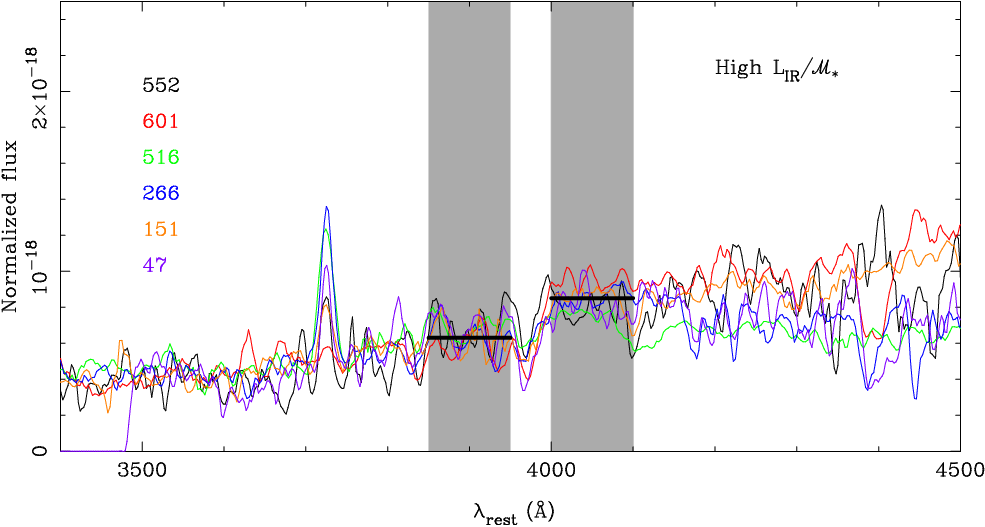}
 \includegraphics[width=4.5cm,angle=-90]{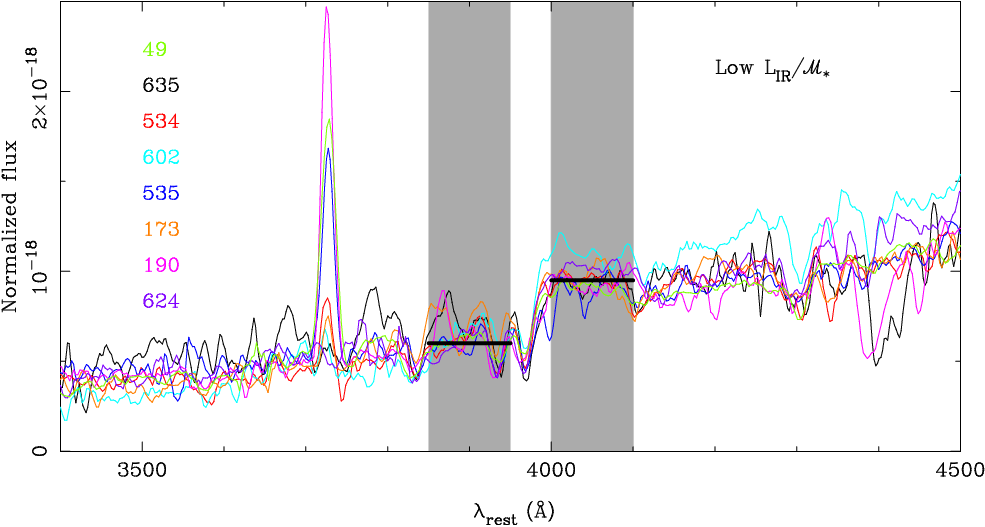}

%for preprint
% \includegraphics[width=6cm,angle=-90]{figure4a.ps}
% \includegraphics[width=6cm,angle=-90]{figure4b.ps}

\caption{Normalized rest-frame spectra (from Szokoly et al. 2004) showing the
  spectral region of the [O\,{\sc ii}]$\lambda$3727 
line and the 4000\AA \, break for AGN in our sample 
at $0.5 < z < 0.8$. The
  shaded regions indicate the band-passes defined by Balogh et al. (1999) to
  measure the $D_n(4000)$ index, and used by Kauffmann et al. (2003a,b) to
  derive the star formation histories of local AGN. 
The thick lines in the band-passes are plotted to guide the eye
  and correspond to the approximate average continuum level. 
The top and bottom panels show AGN with high and low 
$L_{\rm IR}/\mathcal{M}_*$ ratios (galaxies from the upper left and upper
  right panels 
of Fig.~1),  respectively. } 
\end{figure}

Since most of the optically-dull AGN with low absorptions have only 
modest X-ray
luminosities ($L_{\rm 2-10 keV} < 2-3 \times 10^{42}\,{\rm erg \ s}^{-1}$, 
see Fig.~2), 
we consider the possibility that they do not contain an AGN, and thus that 
their X-ray luminosities could be produced by 
star formation. To do so, we can compare our total
UV+IR SFRs with the SFRs predicted from their hard X-ray luminosities using
the relation inferred by Ranalli, Comastri, \& Setti (2003). In all but one
galaxy (Szokoly 577, which is one of the galaxies with high 
$L_{\rm IR}/\mathcal{M}_*$ ratios) the X-ray 
based SFRs would be between 3 and 10 times higher
than our inferred SFR$_{\rm UV+IR}$, indicating that in these galaxies most of
the hard X-ray emission is not produced by star formation, and that indeed
they contain (a low luminosity) AGN.

\subsection{The average 
stellar ages of the host galaxies}

In Fig.~1 the observed SEDs (normalized at $\lambda_{\rm rest} = 1-2\,\mu$m)
are shown for the two redshift bins, and are sorted according to their fitted 
$L_{\rm IR}/\mathcal{M}_*$ ratios. If the AGN contribution to the MIR emission
is small, these ratios are a good proxy for 
the specific SFRs (i.e., star formation rate per unit stellar mass). 
%We also plot in the low redshift bin 
%the average of the models fitted for galaxies with high and 
%low $L_{\rm IR}/\mathcal{M}_*$. 
From the modelling of the stellar SEDs (see \S3) we find that most galaxies
with high $L_{\rm IR}/\mathcal{M}_*$ ratios 
are fitted with younger models than those
with low $L_{\rm IR}/\mathcal{M}_*$ ratios.

An independent way to estimate the average ages of the host galaxy stellar
populations is to measure the 
4000-\AA \, break (see Kauffmann et al. 2003a,b and references therein).
The optical spectra (from Szokoly et al. 2004) of some $0.5<z<0.8$ 
\footnote{At higher
redshifts this feature is too close to the edge of their 
optical spectra.}  AGN with high and low $L_{\rm IR}/\mathcal{M}_*$ 
ratios (as in Fig.~1) are shown in Figure~4.
The  average $D_n(4000)$ values for the AGN in Figure~4 with high and low 
$L_{\rm IR}/\mathcal{M}_*$ ratios are $\sim 1.4$ and $\sim 1.7$, respectively. 
Thus, there is some marginal evidence for the 
AGN host galaxies with the lowest $L_{\rm IR}/\mathcal{M}_*$ ratios 
 to show larger 4000-\AA \,
  breaks (i.e., older stellar populations) than 
galaxies with high $L_{\rm IR}/\mathcal{M}_*$ ratios. 
These $D_n(4000)$ values are similar to  
those of high-$z$  AGN identified by Kriek et al. (2007). 

The measured  $D_n(4000)$ values for our sample of AGN 
indicate relatively young 
ages, of  between 0.8 and 
2.5\,Gyr  for an instantaneous burst 
(using figure~2 of Kauffmann et al. 2003a). These ages, as well as  the ages
  of local universe type-2 AGN (Kauffmann et al. (2003b) and 
intermediate-$z$ type 1 AGN (S\'anchez et al. 2004)   are younger
than those of local quiescent massive galaxies, but consistent with the ages of
IRAC-selected galaxies of similar masses (see P\'erez-Gonz\'alez et al. 2008).

Those galaxies in our sample (about one-third) with high 
$L_{\rm IR}/\mathcal{M}_*$ ratios (left panels
of Fig.~1) tend to show an excess of 
$\lambda_{\rm rest} \sim 3-5 \,\mu$m emission over the expectations
of the stellar emission for an evolved stellar population. Although in most
cases these MIR excesses can be accounted for by the gas + stellar 
emission associated
with a young stellar population, we cannot
rule out some contribution from warm dust emission
associated with the putative AGN. 
By our selection criteria (i.e., prominent $1.6\,\mu$m 
stellar bump), the MIR excesses in
our galaxies occur at $\lambda_{\rm rest} > 3\,\mu$m,
unlike the MIR excess galaxies of Daddi et al. (2007) and IR power-law
galaxies (Alonso-Herrero et al. 2006; Donley et al. 2007), 
where the MIR excesses appear at $\lambda_{\rm rest} \sim 1.6\,\mu$m.  
However, a large fraction (two-thirds) of the sample 
show no MIR excesses out to 
$\lambda_{\rm rest} \sim 5\,\mu$m, 
with their MIR emission being entirely consistent with that
of an evolved stellar populations. Thus in these galaxies the 
AGN emission appears completely
buried out to $\lambda_{\rm rest} < 5\,\mu$m and possibly longer wavelengths.

%Although solely based on the shape of the MIR continuum  it is
%difficult to distinguish between dust heated by star formation and dust heated
%by an AGN, we find that optically-dull AGN with rest-frame 
%$3-5\,\mu$m excesses do not
%show $f_\nu(70\,\mu{\rm m})/f_\nu(24\,\mu{\rm m})$ ratios typical of type 1
%AGN (i.e., unobscured view of the AGN), indicating that even in optically-dull 
%AGN with MIR  excesses the AGN emission 
%probably does not dominate at rest-frame wavelengths longer than $\sim
%30-40\,\mu$m   (see also Daddi et al. 2007).

\section{Host galaxy properties}

\subsection{Stellar Masses}

Figure~5 shows the stellar masses versus redshift for our sample of
AGN compared with the distribution of stellar masses for the
IRAC-selected sample of galaxies of P\'erez-Gonz\'alez et al. (2008). 
At the redshifts probed here the IRAC-selected comparison sample is essentially a
stellar mass selected sample. 
Clearly
 (X-ray identified) AGN reside in galaxies with 
a range of about an order of magnitude in mass, including some among the 
most massive at these intermediate redshifts. 
The characteristic stellar masses (measured as 
the median of the
distributions) are $7.8 \times 10^{10}\,{\rm M}_\odot$ (36 galaxies) 
at $0.5<z<0.8$ (median
redshift of 0.67) and $1.2 \times 10^{11}\,{\rm M}_\odot$ (22 galaxies) 
at $0.8<z<1.4$ (median redshift of 1.07). 
We also plot in Fig.~5 as solid lines the redshift evolution of  the quenching
mass (mass above which, star formation  should be mostly suppressed)
inferred by Bundy et al. (2006) using two different methods. The fraction of
AGN above the line  is small perhaps indicating that star formation has not
been fully suppressed yet in these galaxies (but see also \S5.2).

\begin{figure}
%\vspace{1cm}

 \includegraphics[width=8.5cm]{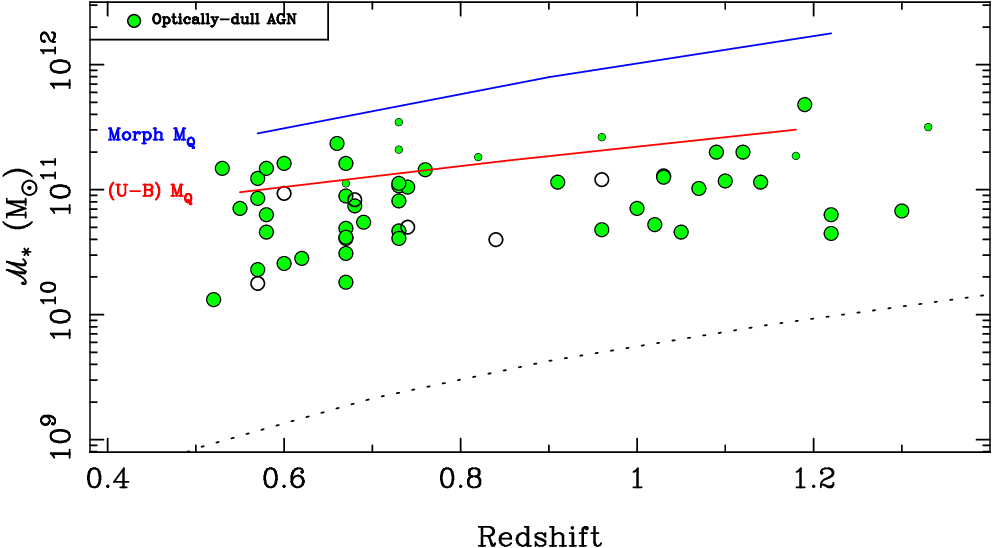}
 \includegraphics[width=8.5cm]{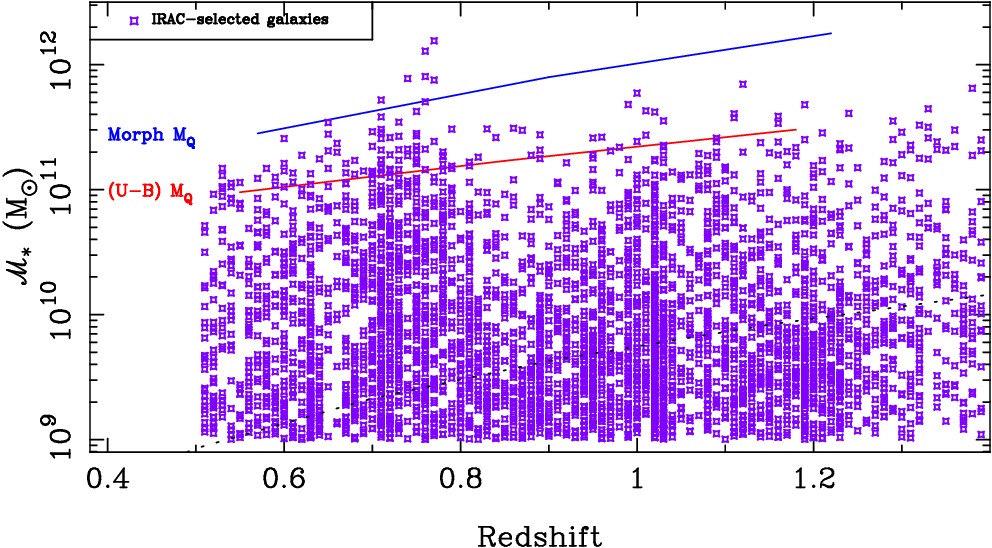}

%for preprint format
% \includegraphics[width=12cm]{figure5a.ps}
% \includegraphics[width=12cm]{figure5b.ps}
\caption{{\it Upper panel:} 
Redshift evolution of the stellar mass of our sample of AGN 
in the  CDF-S (all circles). We marked with small filled circles those  optically-dull AGN with 
$\mathcal{M}_*>10^{11}\,{\rm
    M}_\odot$ and specific SFRs similar or 
 below the median specific SFRs for IRAC-selected galaxies (see more details
 in \S5.2). 
The open circles are the stellar masses for  the six optically-active AGN in
our sample with stellar SEDs. 
The dotted line indicates the completeness limit of the sample of 
P\'erez-Gonz\'alez et al. (2008) for a maximally-old passively evolving
galaxy. The two  
solid lines are two empirical determinations (color and morphology) of the 
redshift evolution of the quenching mass (converted to a
Salpeter IMF) 
of Bundy et al. (2006). The solid 
lines reflect two different criteria used by Bundy
et al. (2006) to 
evaluate star formation: color and morphological type.  
{\it Lower panel:}
  Redshift evolution of the stellar mass of IRAC-selected galaxies
  (P\'erez-Gonz\'alez et al. 2008) in the same field as our sample of 
CDF-S AGN. The IRAC includes all the AGN plotted
  in the upper panel. Only galaxies with $\mathcal{M}_*>10^{9}\,{\rm
    M}_\odot$ are plotted in this comparison.} 
\end{figure}

It is important to stress that
the  AGN studied here comprise $\sim 50\%$ of the X-ray selected 
AGN population at $0.5 < z < 1.4$. 
A pressing  question is whether the derived stellar masses 
are representative of the overall population of X-ray selected AGN. 
One possibility is that  AGN with stellar-dominated SEDs (mostly
optically-dull AGN) might be hosted by  
massive galaxies (e.g., Moran et al. 2002; 
Severgnini et al. 2003;  Caccianiga et
al. 2007) causing the AGN emission lines to be buried by the galaxy emission. 
To test this possibiliy, in Fig.~6, which is similar to Fig.~2, the 
 AGN in our sample are sorted according to their stellar masses. There is no
 clear tendency for the most X-ray luminous sources to be hosted by the most
 massive galaxies.  As demonstrated by Rigby et al. (2006) 
extinction on large scales produced by the host galaxy might also be 
responsible for 
hiding the AGN lines in these galaxies.

Ideally we would like to estimate 
the stellar masses for all  optically-active AGN at intermediate redshifts,
but this   becomes increasingly 
more  uncertain (see P\'erez-Gonz\'alez et al. 2008), 
as for more luminous X-ray sources  
the AGN emission in the optical-NIR becomes more dominant (e.g., 
Barmby et al. 2006; Polletta et al. 2006, 2007, Donley et
al. 2007; see also next section). The masses of 
the six optically-active AGN in our
sample do not appear to be fundamentally different from
optically-dull AGN (see Fig.~5), although the number statistics is
very small.  

Another way to determine if the stellar masses of
optically-dull AGN are representative is to compare their absolute magnitudes
with those of optically-active  AGN. If both types of AGN reside in 
similar systems, then 
optically-active AGN should be more luminous  in the optical and near-IR
because they should have the contributions from  the host galaxy and the AGN. 
Rigby et al.
(2006, their figure~8) already made this comparison, and found that 
BLAGN with  $L_{\rm X} > 10^{43}\,{\rm erg\,s}^{-1}$ tend to
show brighter optical magnitudes than optically-dull AGN of similar
luminosities. Again the number statistics are  small, but both comparisons seem
to suggest that the stellar masses of optically-dull AGN are representative of
the whole X-ray selected AGN population at these redshifts.

\begin{figure}
 \includegraphics[width=8cm,angle=-90]{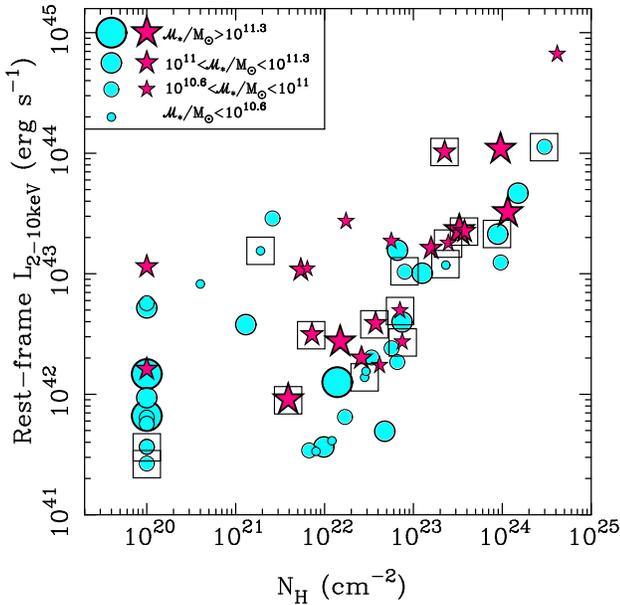}

%for preprint format
%\hspace{2cm}
% \includegraphics[width=10cm,angle=-90]{figure6.ps}
\caption{This figure is similar to Figure~2, but we only plot the 
the  AGN selected for this study, coded according to their derived stellar 
masses for the two redshift bins considered here 
($0.5 < z < 0.8$ circles and $0.8<z<1.4$ stars). We also marked with squares 
those optically-dull AGN with $\lambda_{\rm rest} \sim 3-5\,\mu$m excesses
over pure stellar emission from an old stellar population 
(left panels of Figure~1).}
\end{figure}

Optically selected type-2 AGN at
$z<0.3$ are also found to reside in the most massive galaxies. 
Kauffmann et al. (2003b)
inferred a characteristic stellar mass of  
$\sim 4-5 \times 10^{10}\,{\rm M}_\odot$ (for a Salpeter IMF)
for bright type 2 AGN (see also Heckman et al. 2004). 
At high redshift ($z\sim 2$) X-ray selected AGN appear to be hosted by even
more massive galaxies ($\mathcal{M}_* \sim 1-2 \times 
10^{11}\,{\rm M}_\odot$), although the mass estimates have only 
been done for a few galaxies  (see 
Borys et al. 2005;  Daddi et al. 2007, but also caveats discussed by
 Alexander et  al. 2007). Similarly Kriek et
al. (2007) find that $z \sim 2.3$ 
AGN identified in $K$-band selected galaxies are hosted
by very massive galaxies, but as they point out it is likely that their
results are biased towards the most massive objects.  
The stellar masses of AGN at $ z \sim 0.7$ and 
$z \sim 1.1$ are thus intermediate between those of local AGN and 
high-$z$ AGN.

Since the stellar masses of our sample of AGN appear to be representative
of the whole population of X-ray selected AGN (see Fig.~5) then 
we infer that approximately 25\% of massive galaxies
($\mathcal{M}_*>10^{11}\,{\rm M}_\odot$) 
at the redshifts probed here 
contain an X-ray identified AGN. We also took into account 
AGN not included in this study, assuming that they have stellar masses
similar to the AGN with stellar dominated SEDs. 
This fraction
is similar for the two redshift bins considered here, and consistent with
the AGN fraction in massive galaxies at higher redshifts ($z> 1 $, Alexander et
al. 2005; Daddi et al. 2005; Papovich 
et al. 2006; Kriek et al. 2007). Our 
estimated AGN fraction in intermediate-$z$ 
massive galaxies is only a
lower limit because presumably most Compton-thick AGN are missed by 
current X-ray surveys. At $z\sim 2$ the fraction of AGN in massive galaxies
appears to be as high as $50-60\%$ (Daddi et al. 2007) consistent with the
view that the massive black hole growth  was
higher when the universe was younger.

\subsection{Star-formation activity}

\begin{figure}

 \includegraphics[width=7.9cm]{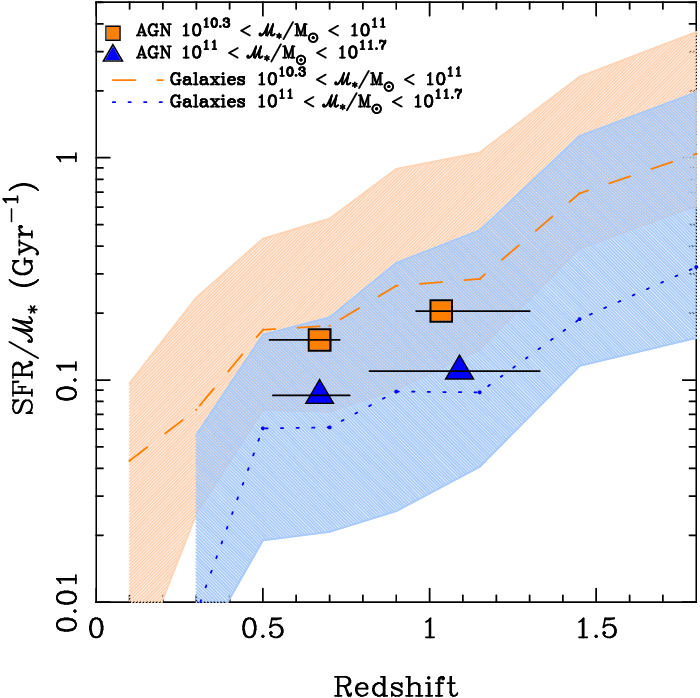}

%for preprint format

%\hspace{2cm}
% \includegraphics[width=10cm]{figure7.ps}
\caption{Redshift evolution of the specific SFR for IRAC-selected 
galaxies in a number of
  cosmological fields. This figure is an updated version (it now includes
  additional 
data for the Extended Groth Strip, P\'erez-Gonz\'alez et. al. 2008, in
  preparation) of 
figure~10 by P\'erez-Gonz\'alez et al. (2008), but here we only show the 
mass intervals of interest for our sample of AGN. 
The dashed and dotted lines are the median specific SFR for 
 the $2\times 10^{10} < \mathcal{M}_*/{\rm M}_\odot < 10^{11}$ 
and $10^{11} < \mathcal{M}_*/{\rm M}_\odot 
< 5 \times 10^{11}$ mass ranges, respectively. 
We also show the quartiles of the distribution of specific SFRs of 
IRAC-selected
  galaxies for each of the two mass ranges as the 
shaded regions. The median specific SFRs for AGN
  are the filled squares and triangles plotted at the median redshift. The
  horizontal bars represent the ranges of redshifts. We  excluded
 AGN with MIR excesses or classified as optically-active. }
 \end{figure}

 In this section we quantify the star
formation activity of AGN with stellar-dominated SEDs in relation to IRAC-selected
galaxies of similar stellar masses and at similar redshifts. 
We use the specific SFRs as an indicator of the star formation activity rather
than the absolute SFRs, as the specific 
SFR measures the rate at which 
new stars add to the assembled mass of the galaxy (Brinchmann \& Ellis
2000). Moreover, galaxies show distinct specific SFRs depending on
their stellar masses and redshifts (e.g., 
Brinchmann et al. 2004; 
Zheng et al. 2007; P\'erez-Gonz\'alez et al. 2008), so the comparison needs to
be made taking this into account.

Figure~7 compares the median specific SFRs for
intermediate-$z$ AGN   with those of the
IRAC-selected sample of P\'erez-Gonz\'alez et al. (2008).
In this comparison we excluded MIR-excess galaxies  and the six
optically-active AGN, as 
they may have an important 
AGN contribution to their
observed MIR emission which would cause us to 
overestimate their IR-based SFRs.  
The comparison between AGN and IRAC-selected galaxies 
is done for the two 
relevant stellar mass ranges of the AGN hosts (see Figure~5):
$2 \times 10^{10}\,{\rm M}_\odot < \mathcal{M}_* < 10^{11}\,{\rm M}_\odot$
and $10^{11}\,{\rm M}_\odot < \mathcal{M}_* < 5 \times 10^{11}\,{\rm
  M}_\odot$.  

As can be seen from Figure~7 the AGN specific SFRs 
(at both redshift and mass intervals) do not appear to be fundamentally
different from those of IRAC-selected galaxies. 
Moreover, Zheng et al. (2007) argued against AGN feedback as the main process for 
quenching star formation. Their model predicted 
that AGN feedback would be more effective for more massive galaxies (see also Croton et
al. 2006). This would imply a
more rapid decline of the specific SFR for massive galaxies ($>10^{11}\,{\rm
  M}_\odot$) than less massive galaxies (see also Springel et al. 2005), which
is observed neither for the Zheng et al. (2007) sample nor for the
IRAC-selected sample of P\'erez-Gonz\'alez et al. (2008, see Fig.~7). 

In contrast,  Kriek et
al. (2007) found evidence for a relation between the suppression of star
formation and the AGN phase for $K$-band selected galaxies. Kriek et
al. (2007) however pointed out that their AGN
sample may be biased toward quiescent galaxies where AGN are easier to
identify.  In fact, if we consider their two samples of galaxies (UV and
$K$-band selected) then the AGN have specific SFRs within the range observed
for their non-AGN. 
In the local universe AGN tend to be hosted in
massive galaxies with younger stellar ages than non-AGN  of similar
morphological types (early-type) and stellar masses (Kauffmann et al. 2003a, 2007). This was 
interpreted  as 
evidence that enhanced star formation is a requisite for
feeding the AGN. At $z\sim 1$ galaxies (presumably both AGN and non-AGN) with 
$\mathcal{M}_*  \sim 10^{10}- 5 \times 10^{11}\,{\rm M}_\odot$ are still being
assembled (see P\'erez-Gonz\'alez et al. 2008), so
perhaps it 
is not surprising that their star formation rates are not 
significantly different.

\begin{figure}

 \includegraphics[width=8cm,angle=-90]{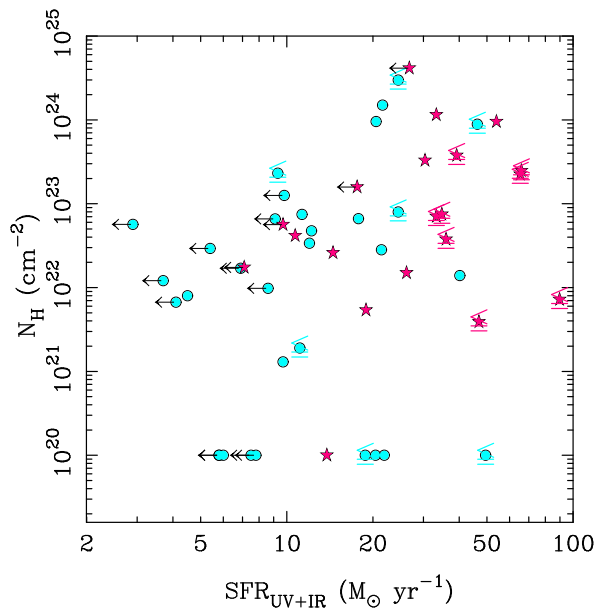}

%for preprint format

%\hspace{2cm}
% \includegraphics[width=10cm,angle=-90]{figure8.ps}
\caption{Total SFR versus the X-ray column densities for the sample of
  optically-dull AGN. Symbols are as in Figure~6 for the two redshift 
bins. The SFRs of galaxies with $24\,\mu$m flux densisites below 
  $80\,\mu$Jy are marked as upper limits.  Galaxies with rest-frame
  $3-5\,\mu$m excesses (left panels of Figure~1) are shown
 with the ``less than or equal'' symbols, as their MIR emission might include
  some  AGN contribution.}
 \end{figure}

\subsubsection{Caveats}

There are a number of caveats when trying assess the star formation activity
of our sample of AGN. As discussed in previous sections, there is the
exact AGN contribution to the MIR emission, although it is found generally not
to be large (\S4.1). Also the AGN in the $0.8 < z < 1.4$ redshift range  tend to be
more luminous in X-rays so the AGN become more apparent. Even though we excluded galaxies with MIR
excesses (as we cannot rule out that they might be due to dust heated by the
AGN), the specific SFRs are formally upper limits. 
Another largely unknown effect in the comparison between AGN and IRAC-selected
galaxies is the 
possible 'contamination' by X-ray identified AGN as well as obscured AGN 
in the most massive ($\mathcal{M}_*  > 10^{11}\,{\rm M}_\odot$) systems 
(see \S5.1 and Daddi et al. 2007; Kriek et al. 2007)
in the comparison sample.

Another concern is the possibility that the 
obscuring material, which may be responsible in part for the optical dullness,  
be associated with star formation activity 
within the host galaxy (e.g., Ballantyne, Everett, \& Murray 2006; 
Mart\'{\i}nez-Sansigre et al. 2006). This would bias our sample towards
star-forming galaxies when compared to the most optically-active AGN not
included in our sample.  
Fig.~8 shows a comparison between the derived UV+IR SFRs and the X-ray
absorption for our sample of AGN. This figure suggests a weak, if any, connection
between star formation and obscuration. It is also clear that a number of
highly obscured AGN ($N_{\rm H}>10^{22}\,{\rm cm}^{-2}$) do not present high
SFRs, and in these cases the obscuration is probably not associated with
extranuclear dust  in the host galaxy.

Summarizing, at intermediate-$z$ we do not find strong 
evidence for either highly suppressed star formation activity or increased
star formation activity in AGN when compared to IRAC-selected 
galaxies of similar stellar masses. We only observe that the most massive 
galaxies with low
specific SFRs (see Fig.~5) follow the 
redshift evolution of  the quenching mass 
inferred by Bundy et al. (2006). This may only indicate that the most massive
systems hosting an AGN are close to being fully assembled at these redshifts.

\section{Black hole masses and accretion rates of typical AGN}
The relation between the 
black hole mass and the bulge luminosity and in particular the bulge 
stellar mass is now well estalished 
in the local universe 
(e.g., Merritt \& Ferrarese 2001; Marconi \& Hunt 2003). This
relationship appears to hold out to $z\sim 1$ (Peng et al. 2006), indicating
that massive bulges were fully assembled at this redshift (see e.g.,
Glazebrook et al. 2004; Cimatti et al. 2004; Papovich
et al. 2006; 
P\'erez-Gonz\'alez et al. 2008).  Since the AGN activity
out to $z \sim 1.3$ seems to be associated with bulge-dominated galaxies (see
S\'anchez et al. 2004; Grogin et al. 2005; 
Pierce et al. 2007) we can use the local relationship and 
our stellar masses to estimate the black hole masses
of intermediate-$z$ AGN.

Using the Marconi \& Hunt (2003)  relation and assuming that 
$\mathcal{M}_{\rm bulge}
\approx \mathcal{M}_*$, we find  
 black hole masses
for typical AGN of approximately 
between $4 \times 10^7$ and $10^9\,{\rm M}_\odot$, with a median value of 
$\mathcal{M}_{\rm BH} \sim 2 \times 10^{8}\,{\rm M}_\odot$. These are in good
agreement with the estimates of Babi\'c et al. (2007) for CDF-S $z \sim 0.7$ 
AGN based on stellar masses and stellar velocity dispersions. 
The resulting $\mathcal{M}_{\rm BH}$ of optically-dull AGN 
are plotted in Figure~9 against the
absorption-corrected rest-frame  
hard X-ray luminosities\footnote{The sizes of the symbols are proportional to
  their rest-frame 
monochromatic $24\,\mu$m luminosities. If most of the MIR emission was
produced by dust heated by an AGN, one would expect a proportionality between
the absorption corrected X-ray luminosities (a proxy for the AGN luminosity) 
and the MIR luminosities, which is
not observed in general for optically-dull AGN (see also Rigby et al. 2006). 
This again  suggests that a
large fraction of the MIR emission relative to the AGN luminosity 
could be due to star
formation.}.

\begin{figure}

\includegraphics[width=7.9cm,angle=-90]{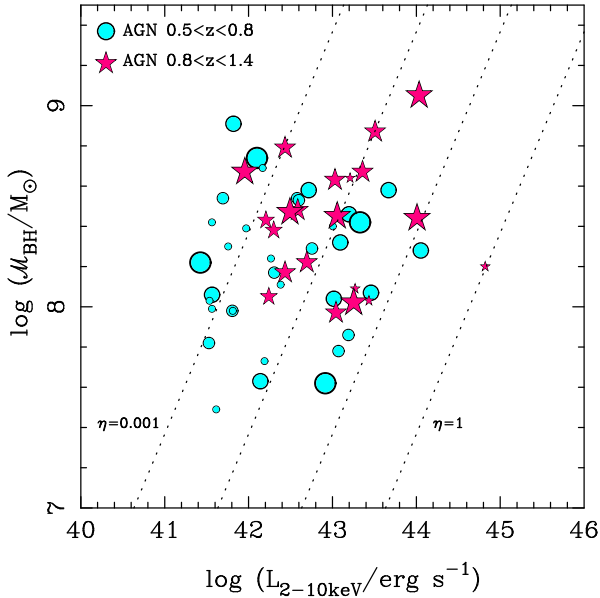}

%for preprint format

%\hspace{2cm}
% \includegraphics[width=10cm,angle=-90]{figure9.ps}
\caption{Black hole masses (derived from the 
$\mathcal{M}_{\rm BH}/
\mathcal{M}_{\rm bulge}$ relation of Marconi \& Hunt 2003) 
versus absorption-corrected rest-frame 
hard X-ray luminosities. The circles are AGN at $0.5 < z < 0.8$ and the
star symbols are AGN at $0.8 < z < 1.4$. 
The sizes of the symbols are proportional to their 
rest-frame $24\,\mu$m luminosities. 
For both redshift bins, the smallest
symbols are  galaxies not detected at $24\,\mu$m or flux densities below 
$80\,\mu$Jy. 
The dotted lines indicate from right to left Eddington ratios of  $\eta =1, 0.1, 0.01$ and 0.001, assuming 
a bolometric correction of 
$L_{\rm bol}/L_{\rm X} \sim 30$ as derived by  
Elvis et al. (1994). }
 \end{figure}

In Figure~9
we also show the Eddington ratios ($\eta=L_{\rm bol}/L_{\rm Edd}$) 
calculated using the bolometric
corrections ($L_{\rm bol}/L_{\rm X} \sim 30$) derived for PG quasars by 
Elvis et al. (1994). Optically-dull AGN at $0.5<z<1.4$ with X-ray
luminosities above $10^{43}\,{\rm erg\,s}^{-1}$ have Eddington
ratios close  to broad-line AGN ($\sim 0.16$, Barger et
al. 2005). However for X-ray luminosities below 
$10^{43}\,{\rm erg\,s}^{-1}$ (comparable to local Seyfert galaxies) 
the Eddington ratios are much lower ($\eta 
\sim 0.01-0.001$, similar to the findings
of Babi\'c et al. 2007).

One possibility for the low Eddington ratios is that we were 
overestimating significantly 
the black hole masses, for instance,  if these galaxies were not
bulge-dominated. However, the corrections to the bulge masses (and thus black
hole 
masses) would be of the order of $\sim
2$ to $\sim 5-6$ for S0/a and Sc types, respectively (see e.g., Marconi et 
al. 2004; Dong \& De Robertis 2006), assuming that the NIR luminosity traces
the stellar mass.  Recently Ballo et al. (2007) derived bulge magnitudes for
X-ray sources in the CDF-S and found that in general in the ACS 
$z$-band the bulge
to total luminosity ratios are of beween 0.4 and 1. For the two souces in our
sample in common theirs these ratios are $0.6-0.8$. From this, the black hole
masses quoted here should be taken as upper limits and Eddington
ratios  lower limits, although the corrections are likely to be
 relatively small.

The AGN hosts in our sample   
follow very well the redshift evolution of the 
quasar host mass derived by
Hopkins et al. (2007) from the quasar optical luminosity function. 
This is interesting because our AGN are not X-ray quasars (i.e., 
$L_{\rm X}<10^{44}\,{\rm erg\,s}^{-1}$,  see Fig.~2). 
This is also  indirect evidence   
that the accretion rates of optically-dull AGN are lower than those of bright
quasars. In \S5.1 on the other hand we found that 
the characteristic stellar masses
of intermediate-$z$ are between those of local AGN and high-$z$ ($ z >2$)
AGN. This would 
appear to support the interpretation 
of Heckman et al.
(2004) for the AGN 'downsizing' phenomenon. In this scenario, the fact that 
the space density of low luminosity  AGN peaks at lower redshifts than more
luminous AGN (see Ueda et al. 2003) is explained by a decrease of the
characteristic mass of actively accreting black holes (and thus, stellar mass
of the hots) rather than a
decreasing accretion rate (see also Barger et al. 2005). 
Since it is these low X-ray luminosity  AGN that dominate
the X-ray background at the redshifts considered here, the AGN 'downsizing' is
not only due to a decrease in the characteristic stellar mass (see \S5.1), but
also lower Eddington ratios. 

%\subsection{Is star formation responsible (in part) for obscuring
%  optically-dull AGN?}

\section{Summary and Conclusions}

We studied a sample of 58 X-ray selected AGN at
intermediate redshifts ($z \sim 0.5-1.4$) 
in the GOODS portion of the CDF-S. The AGN were selected such that their
rest-frame UV to NIR SEDs were dominated by stellar emission, and in
particular, showed a prominent $1.6\,\mu$m bump. This selection minimized the
AGN contamination which is essential for studying the properties of their host galaxies. 
Since AGN with stellar dominated SEDs comprise approximately 50\% of the
population of X-ray selected AGN, they are a cosmologically important class of
AGN. We fitted their rest-frame UV through MIR SEDs using stellar and dust models to
derive the stellar masses as well as the total (UV+IR) SFRs. 

As previously discussed by other works (e.g., Severgnini et al. 2003; Rigby
et al. 2006; Caccianiga et al. 2007), the optical dullness might be due to
various causes. We find that dilution by a massive host galaxy is only in part
responsible, as the mass of the host
galaxy is independent of the X-ray
luminosity and absorption of the AGN. Extinction on large scales may also play
a role, and presumably is relatively prevalent in the lower-mass
optically dull AGN hosts.  
From this and the derived stellar
masses of a few optically-active AGN with stellar SEDs we conclude that the
derived stellar masses of our sample of AGN are representative of the entire
X-ray selected AGN population at these redshifts. 

About one-third of our AGN show 
$\lambda_{\rm rest} \sim 3-5\,\mu$m excesses above the expected stellar
emission from an old stellar population 
together with high $L_{\rm IR}/\mathcal{M}_*$ ratios. Although these MIR
excesses
 could be interpreted as evidence for the putative AGN (i.e., hot dust), 
these galaxies tend to have smaller 4000\AA-breaks. That is, galaxies with MIR
excesses have younger average
stellar populations (an indication of recent or on-going star formation) 
than galaxies with low $L_{\rm IR}/\mathcal{M}_*$
ratios. This may indicate that  AGN emission is 
responsible for most of the  MIR emission only in some galaxies.  For the
rest of the sample there is no evidence for AGN emission out to approximately 
$\lambda_{\rm rest} \sim 5\,\mu$m, and their MIR emission is likely to be
produced mostly by star formation.

X-ray identified AGN are found to reside in galaxies with a range of stellar
masses ($\mathcal{M}_* 
\sim 2 \times 10^{10}-5\times 10^{11}\,{\rm M}_{\odot}$, for a Salpeter 
IMF), including the most massive
galaxies at intermediate redshifts. 
We infer characteristic (median) stellar masses  of 
$\mathcal{M}_* \sim  7.8 \times 10^{10}\,{\rm M}_\odot$ 
 and $\mathcal{M}_*  \sim 1.2 \times 10^{11}\,{\rm M}_\odot$ 
at  median redshifts of 0.67 
and 1.07, respectively. These stellar masses 
are intermediate between those of local type 2 AGN ($\sim 4
\times 10^{10}\,{\rm M}_\odot$,
Kauffmann et al. 2003b) and
high-$z$ AGN ($\sim 1-3 \times 10^{11}\,{\rm M}_\odot$, Kriek et al. 2007;
Daddi et al. 2007; Alexander et al. 2007).  
From the comparison with the
IRAC-selected sample of P\'erez-Gonz\'alez et al. (2008), 
we find that approximately 25\% of massive
($\mathcal{M}_* > 10^{11}\,{\rm M}_\odot$) galaxies at these redshifts contain
an X-ray identified AGN. 

Using the local relation between 
$\mathcal{M}_{\rm bulge}$ 
and $\mathcal{M}_{\rm BH}$ of Marconi \& Hunt (2003) the inferred 
black hole masses  ($\mathcal{M}_{\rm BH} \sim 2 \times
10^{8}\,{\rm M}_\odot$) are similar to
those of broad-line AGN although with lower Eddington ratios ($\eta \sim
0.01-0.001$) than luminous quasars. 
Both findings suggest that, at least at intermediate redshifts, 
the cosmic AGN 'downsizing' is probably due not only to a 
decrease in the characteristic stellar mass of the host galaxy (as proposed by
Heckman et al. 2004), but also to less
efficient accretion, as also found by Babi\'c et al. (2007). 

Finally, we do not find strong evidence in AGN host galaxies 
for either highly suppressed star
formation (expected if AGN played a role in quenching star formation) or
intense star formation when compared to IRAC-selected galaxies of similar
stellar masses and redshifts. This can be understood if we take into account
that the host galaxies of AGN (and non-AGN of similar mass) 
are still being assembled at the redshifts probed 
here.  The main caveats regarding this conclusion 
are the possible influence on it of the AGN contribution to the observed
MIR  emission (although it is found generally not to be large), 
and the possible bias 
 towards intense star formation which could be in part responsible for
 obscuring the AGN in these objects. 

$\,$

The authors would like to thank A. Marconi, 
P. Hopkins, and M. Kriek for interesting
 discussions. The authors also thank an anonymous referee for 
 useful suggestions that helped improve the paper.
 This work was supported by NASA through contract 1255094 issued by
 JPL/California Institute of Technology.  AA-H acknowledges support from 
the Spanish Plan Nacional del Espacio under grant ESP2005-01480.
PGP-G acknowledges support from the Ram\'on y Cajal Fellowship Program
financed by the Spanish Government, and from the Spanish Programa
Nacional de Astronom\'{\i}a y Astrof\'{\i}sica under grant AYA
2006--02358.
This research has made use of the NASA/IPAC Extragalactic Database (NED),
 which is operated by the Jet Propulsion Laboratory, California Institute of
 Technology, under contract with the National Aeronautics and Space
 Administration.


\begin{thebibliography}{99}

\bibitem{alex} Alexander, D. M.  et al. 2003, AJ, 126, 539
\bibitem{alex05}Alexander, D. M., Smail, I., Bauer, F. E., 
Chapman, S. C., Blain, A. W., Brandt, W. N., \& Ivison, R. J. 
2005, Nature, 434, 738
\bibitem{alex07}Alexander, D. M. et al. 2007, ApJ, submitted
%\bibitem{aah00}  Alonso-Herrero, A. et al. 2000, ApJ, 530, 688
%\bibitem{aah04} Alonso-Herrero, A. et al. 2004, ApJS, 155, 154
\bibitem{aah06}Alonso-Herrero, A. et al. 2006, ApJ, 640, 167
\bibitem{babic}Babi\'c, A., Miller, L., Jarvis, M. J., 
Turner, T. J., Alexander, D. M., \& Croom, S. M. et al. 2007, A\&A, submitted
\bibitem{ballantyne}Ballantyne, D., Everett. J. E., \& Murray, N. 
2006, ApJ, 639, 740
\bibitem{ballo}Ballo, L. et al. 2007, ApJ, 667, 97
\bibitem{balogh}Balogh, M. L. et al. 1999, ApJ, 527, 54
\bibitem{barger}Barger, A. J. et al. 2001, AJ, 126, 632 
\bibitem{barger05}Barger, A. J. et al. 2005, AJ, 129, 578 
\bibitem{barmby}Barmby, P. et al. 2006, ApJ, 642, 126
%\bibitem{bauer}Bauer, F. E. et al. 2004, AJ, 128, 2048
\bibitem{bell}Bell, E. F. et al. 2004, ApJ, 608, 752 
\bibitem{bell07}Bell, E. F. et al. 2007, ApJ, 663, 834
\bibitem{borys}Borys, C., Smail, I., Chapman, S. C., Blain, A. W.,
Alexander, D. M., \& Ivison, R. J. 2005, ApJ, 635, 853
\bibitem{brandt}Brandt, W. N. \& Hasinger, G. 2005, ARA\&A, 43, 827
%\bibitem{bohm}B\"ohm, A. et al. 2006, ASP Conference Series, ``Cosmic
%  Frontiers'', astro-ph/0612310
\bibitem{brinchmann}Brinchmann, J. \& Ellis, R. S. 2000, ApJ, 536, L77
\bibitem{brinchmann04}Brinchmann, J. et al. 2004, MNRAS, 351, 1151
\bibitem{bundy}Bundy, K. et al. 2006, ApJ, 651, 120 
\bibitem{caccianiga}Caccianiga, A., Severgnini, P., Della Ceca, R.,
Maccacaro, T., Carrera, F. J., \& Page, M. J. 2007, A\&A, 470, 557 
\bibitem{calzetti}Calzetti, D., Armus, L., Bohlin, R. C., Kinney, A. L., 
Koornneef, J., \& Storchi-Bergmann, T. 2000, ApJ, 533, 682
\bibitem{caputi}Caputi, K. I. et al. 2006, ApJ, 637, 727 
\bibitem{chary}Chary, R., \& Elbaz, D. 2001, ApJ, 556, 562
\bibitem{cimatti}Cimatti, A. et al. 2004, Nature, 230, 184
\bibitem{cohen}Cohen, J. G. 2003, ApJ, 598, 288
\bibitem{croton}Croton, D. J. et al. 2006, MNRAS, 365, 111
\bibitem{daddi05}Daddi, E. et al. 2005, ApJ, 626, 680
\bibitem{daddi}Daddi, E. et al. 2007, ApJ, 670, 173
\bibitem{dong}Dong, X. Y.,  \& De Robertis, M. M. 2006, AJ, 131, 1236
\bibitem{donley}Donley, J. L., Rieke, G. H., P\'erez-Gonz\'alez, P. G., Rigby,
  J. R., \& Alonso-Herrero, A. 2007, ApJ, 660, 167
\bibitem{elvis}Elvis, M. et al. 1994, ApJS, 95, 1
\bibitem{fazio04}Fazio, G. G. et al. 2004, ApJS, 154, 10
\bibitem{fioc}Fioc, M., \& Rocca-Volmerange, B. 1997, A\&A, 326, 950
\bibitem{giacconi}Giacconi, R. et al. 2002, ApJS, 139, 369 
\bibitem{glazebrook}Glazebrook, K. et al. 2004, Nature, 430, 181
\bibitem{grogin}Grogin, N. A. et al. 2005, ApJ, 627, L97
%\bibitem{gilli}Gilli, R. et al. 2000, A\&A, 355, 485
\bibitem{heckman}Heckman, T. M., Kauffmann, G., Brinchmann, J., Charlot, S.,
Tremonti, C., \& White, S. M. D. 2004, MNRAS, 613, 109 
\bibitem{hopkins}Hopkins, P. F., Bundy, K., Hernquist, L., \&
Ellis, R. S. 2007, ApJ, 659, 976 
%\bibitem{juneau}Juneau, S. et al. 2005, ApJ, 619, L135
\bibitem{kauff03a}Kauffmann, G. et al. 2003a, MNRAS, 341, 33
\bibitem{kauff03b}Kauffmann, G. et al. 2003b, MNRAS, 346, 1055
\bibitem{kauff07}Kauffmann, G. et al. 2007, ApJS, in press (astro-ph/0609436) 
\bibitem{kenni98}Kennicutt, R.C. 1998, ARA\&A, 36, 189
\bibitem{kriek}Kriek, M. et al. 2007, ApJ, 669, 776
\bibitem{kuras}Kuraszkiewicz, J. K. et al. 2003, ApJ, 590, 128
\bibitem{lacy}Lacy, M. et al. 2004, ApJS, 154, 166
\bibitem{marconi}Marconi, A., \& Hunt, L. K. 2003, ApJ, 589, L21
\bibitem{marconi04}Marconi, A., Risaliti, G., Gilli, R.,  Hunt, L. K.,
  Maiolino, R., \& Salvati, M. 2004, MNRAS, 351, 169
\bibitem{martin}Martin, D. C. et al. 2007, ApJS, in press (astro-ph/0703281)
\bibitem{martinez}Mart\'{\i}nez-Sansigre, A. et al. 2006, MNRAS, 370, 1479
\bibitem{merritt}Merritt, D., \& Ferrarese, L. 2001, MNRAS, 320, L30
\bibitem{moran}Moran, E. C., Filippenko, A. V., \& Chornock, R. 2002, ApJ, 579, L71
\bibitem{nandra}Nandra, K. et al. 2007, ApJ, 660, L11
\bibitem{norman}Norman, C. et al. 2004, ApJ, 607, 721
\bibitem{papovich}Papovich, C. et al. 2006, ApJ, 640, 92
\bibitem{papovich07}Papovich, C. et al. 2007, ApJ, 668, 45
\bibitem{peng}Peng, C. Y., Impey, C. D., Ho, L. C., Barton, E. J., 
\& Rix, H.-W. 2006, ApJ, 640, 114
\bibitem{perez05}P\'erez-Gonz\'alez, P. G. et al. 2005, ApJ, 630, 82
\bibitem{perez07}P\'erez-Gonz\'alez, P. G. et al. 2008, ApJ, 672, in press
  (astro-ph/0709.1354) 
\bibitem{pierce}Pierce, C. M. et al. 2007, ApJ, 660, L19
\bibitem{polleta}Polletta, M. C. et al. 2006, ApJ, 642, 673
\bibitem{polleta07}Polletta, M. C. et al. 2007, ApJ, 663, 81
\bibitem{ranalli}Ranalli, P., Comastri, A., \& Setti, G. 2003, A\&A, 399, 39
\bibitem{rieke04}Rieke, G. H. et al. 2004, ApJS, 154, 25
\bibitem{rigby04}Rigby, J. R. et al. 2004, ApJS, 154, 160
\bibitem{rigby}Rigby, J. R., Rieke, G. H., Donley, J. L., Alonso-Herrero, A.,
  \& P\'erez-Gonz\'alez, P. G.  2006, ApJ, 645, 115
\bibitem{salim}Salim, S. et al. 2007, ApJS, in press (astro-ph/07043611)
\bibitem{salpeter}Salpeter, E. E. 1955, ApJ, 121, 161
\bibitem{sanchez}S\'anchez, S. F. et al. 2004, ApJ, 614, 586
\bibitem{severgnini}Severgnini, P. et al. 2003, A\&A, 406, 483
\bibitem{springel}Springel, V., Di Matteo, T., \& Hernquist, L. 2005, 
MNRAS, 361, 776
\bibitem{stern}Stern, D. et al. 2005, ApJ, 631, 163
\bibitem{szokoly}Szokoly, G. P. et al. 2004, ApJS, 155, 271
\bibitem{tozzi}Tozzi, P. et al. 2006, A\&A, 451, 457
%\bibitem{treister}Treister, E. \& Urry, C. M. 2006, ApJ, 652, L79 
\bibitem{ueda}Ueda, Y., Akiyama, M., Ohta, K., \& Miyaji, T. 2003, 
ApJ, 598, 886
\bibitem{vanzella}Vanzella, E. et al. 2006, A\&A, 434, 53
\bibitem{zheng}Zheng, W. et al. 2004, ApJS, 155, 73
\bibitem{zhengxz}Zheng, X. Z. et al. 2007, ApJ, 661, L41
\end{thebibliography}
\end{document}